\documentclass[12pt]{iopart}

\usepackage{epsfig,graphics}
\begin{document}

\date{\today}
\title{The estimation of neutrino flux produced by $\mathbf{pep}$ reactions in the Sun }

\author{B.F. Irgaziev}\eads{irgaziev@yahoo.com }
\address{GIK Institute of Engineering Sciences and Technology, Topi,
Pakistan}
\author{V.B. Belyaev}
\address{Joint Institute for Nuclear Research, Dubna, Russia}
\author{Jameel-Un Nabi}
\address{GIK Institute of Engineering Sciences and Technology,
Topi, Pakistan}

\begin{abstract}
The experimental result of the solar neutrino flux at one AU
produced by the $p+p+e \rightarrow d+\nu_e$ reaction
(\textit{pep}) was announced for the first time in 2012 by the
Borexino collaboration. This neutrino flux was significantly
greater than the flux predicted by Bahcall and May, who used two
body approaches for calculation of this reaction. We have used the
three-body model for the proton-proton-electron system in the
continuous spectrum of energy to determine the rate of the
\textit{pep} reaction and have estimated the neutrino flux. Our
result of the neutrino flux is 25-40\% more than the Bachall
\textit{et al.} value and depends on the shape of nucleon-nucleon
($NN$) potential. Moreover, the calculated flux lies within the
confidence interval of the experimental data in the case of pure
attractive potentials as well as potential having repulsion at
small distances between nucleons.
\end{abstract}

\vskip 0.1cm
 \hspace{2.2 cm} (Some figures may appear in color only in the online
journal)

\maketitle

\section{Introduction}
\label{int}

The measurement of the neutrino flux from the Sun gives
possibility to solve several important problems. A careful study
of the solar neutrino flux  penetrating the Earth gives
possibility to understand many characteristics of the neutrino
including the effect of the neutrino oscillation, estimation of
mixing angle and neutrino mass. The neutrino flux along with its
spectral properties give us useful insight to nuclear reactions
which otherwise cannot be observed under terrestrial conditions.
The main reaction called $pp$ reaction going in the interior of
the Sun and determining the luminosity is
\begin{equation}\label{pp}
p+p\rightarrow d+ e^{+} + \nu_e.
\end{equation}
Furthermore, as underlined by John N. Bahcall in his series of
papers and book
\cite{bahcall,astro-ph/0209080,bahcall2004,bahcall-sab1,bahcall-sab2,
www-sns-ias-edu}, solar neutrinos bring the signals from processes
in the core of star,and allow one to make comparison of the data
with solar structure models. Unfortunately it is impossible to
directly measure the neutrino flux from this reaction in the
laboratory. The reaction
\begin{equation}\label{ppe}
p+p+e^{-}\rightarrow d+ \nu_e,
\end{equation}
which is called $pep$ reaction stimulates interest because an
emitted neutrino is monoenergetic even though $pep$ plays no
essential role in hydrogen burning in the Sun. The energy of
neutrino from this reaction is $E_{\nu}=1.44$ MeV. The parameters
of the Standard Solar Model can be estimated from measurement of
the neutrino flux coming from the \textit{pep} reaction.

Bahcall and May \cite{bahcall69} used the two-body approach  for
considering the three-body $pep$ reaction \cite{belyaev}. In the
abstract of Ref. \cite{zub96} the authors presented some
qualitative estimation of the three-body effect for $pep$ reaction
and came to the conclusion that the previous conventional
estimation of the $pep$ solar neutrino flux could have been
underestimated. Therefore the solution of the three-body equation
without applying two-body approximation was required.

The purpose of this work is to treat  the the initial state of the
reaction (\ref{ppe}) as a purely three-body state. It is well
known fact that low energy nucleon-nucleon data is insensitive to
the form of $NN$ potential in the two body case. However, the
three-body system is sensitive to the type of the potential. The
idea behind  this work is to check the sensitivity of neutrino
flux to a  used $NN$ potential, within different Standard Solar
Models, and most important to extend our previous results
\cite{bah13} by using a realistic $NN$ potential.
It took more
than fifty years, after the existence of solar neutrino problem,
for the announcement of the first experimental observation of
process (\ref{ppe}) \cite{borexino}. Further a three-body
description of the initial state in process (\ref{ppe}) with
different types of $NN$ potential was not available in literature.
These were the main motivation for presenting our current
calculation.

In Sec. \ref{input} we briefly introduce our weak Hamiltonian and
the two types of nucleon-nucleon potentials. We determine the
probability and the astrophysical $S_{pep}$ factor of the reaction
(\ref{ppe}) in Section \ref{gen}. The reaction rate and
calculation of solar neutrino flux is presented in Section
\ref{rflux}. We finally conclude and summarize our calculation in
Section \ref{fin}.

In this paper, we applied the same notation for the quantities as
in our previous article \cite{bah13}.

\section{Inputs} \label{input}
Electron capture by the nuclei with emitting electron neutrino is
considered by the application of the  effective Hamiltonian
(nonrelativistic) describing weak interaction presented in Ref.
\cite{prim59}. Taking into account the smallness of the neutrino
energy in the reaction $pep\rightarrow d+\nu_e$ as well as the
ansatz that the transition satisfies the Gamow-Teller selection
rule we reduce the weak transition operator to the form:
\begin{equation}\label{Hweak}
H_w=\tau^{(+)}G_A\sum\limits_{i=1}^2\tau_i^{(-)}\mathbf{\sigma\cdot\sigma_i}\delta{\mathbf{(r-r_i)}},
\end{equation}
where $\mathbf{\sigma}$ and $\mathbf{\sigma_i}$ are spin operators
for the lepton and \textit{i}th nucleon; $\mathbf{r}$ and
$\mathbf{r_i}$ are the space coordinates of the lepton and an
\textit{i}th nucleon; $\tau^{(+)},\,\,\tau^{(-)}_i$ are the
isobaric-spin operators transforming a lepton electron state into
a lepton neutrino state and \textit{i}th nucleon proton state into
an \textit{i}th nucleon neutron state, respectively. We used axial
vector coupling constant $G_A$ equaling to  $G_A/(\hbar
c)^3=-1.454\times 10^{-11}\,\rm{GeV}^{-2}$ \cite{povh}.

In our previous calculation \cite{bah13} we used the simplest
potentials (Gauss and Yukawa types) which are pure attractive in
nature. In this work we apply one more simplest exponential
potential and in addition a realistic $NN$ potential
(Malfliet-Tjon potential \cite{M-T}) having repulsion at small
distance to check sensitivity of the flux from the reaction
(\ref{ppe}) to the shape of potential.

The fitted parameters to these potentials listed below determine
correctly the low energy $NN$ scattering data and binding energy
of the deuteron.

The parameters of the exponential potential
\begin{equation}\label{exp} V^N(r)=-V_{0}\exp\bigl(-r
/R_{N}\bigr)
\end{equation}
describing the low energy $NN$ data at the singlet  ($s$) and
triplet ($t$) states  are
\begin{eqnarray} V_{0}^{s} ={\rm 98.10\; MeV, \; }\,\,\,\, R_{N}^{s} ={\rm 0.744\;
fm}; \nonumber\\
V_{0}^{t} ={\rm 184.08\; MeV,\; \; } R_{N}^{t} ={\rm 0.683
fm}.\nonumber
\end{eqnarray}
These parameters define the following scattering lengths and
effective ranges
\begin{eqnarray}^sa_{pp}&=&-7.874\,\,{\rm{fm}},\,\,\,\,
^sr_{pp}=2.804\,\,{\rm{fm}},\nonumber\\
^ta_{np}&=&\,\,\,\,\,5.403\,\,{\rm{fm}},\,\,\,\,
^tr_{np}=1.716\,\,{\rm{fm}}.
\end{eqnarray}

The M-T potential is
\begin{equation}\label{MT} V^N(r)=\frac{V_{A}}{r/R_A} \exp \bigl(-r
/R_{A}\bigr)+\frac{V_{R}}{r/R_R} \exp \bigl(-r
/R_{R}\bigr).
\end{equation}
Fitting of $NN$ low energy data gives the following parameters of the M-T potentials:
\begin{eqnarray} V_{A} =-{\rm 898.75\; MeV,\; } R_{A} ={\rm 0.617\;
fm},\nonumber\\
 V_{R} ={\rm 4319.85\; MeV,\; \; } R_{R} ={\rm 0.325\,
fm},\nonumber
\end{eqnarray}
 for the singlet state and
\begin{eqnarray}V_{A} ={-\rm 945.50\; MeV,\; } R_{A}^{t} ={\rm 0.645\;
fm},\nonumber\\
 V_{R} ={\rm 4476.845\; MeV,\; \; } R_{R}^{t} ={\rm 0.322\;
fm},\nonumber
\end{eqnarray}
for the triplet state. Here we omitted indexes $s$ and $t$ to simplify entries.
These parameters determine the scattering lengths and
effective ranges:
\begin{eqnarray}^sa_{pp}&=&-7.88\,\,{\rm{fm}},\,\,\,\,
^sr_{pp}=2.69\,\,{\rm{fm}},\nonumber\\
^ta_{np}&=&\,\,\,\,\,5.52\,\,{\rm{fm}},\,\,\,\,
^tr_{np}=1.89\,\,{\rm{fm}}.
\end{eqnarray}
The indexes $s$ and $t$ mean the singlet and triplet state,
respectively.

To find the neutrino flux some Standard Solar Model (SSM) must be
used.  The results of Bahcall \textit{et al.} \cite{bahcall-sab1}
showed that the sensitivity of the flux of neutrino from the $pp$
and $pep$ reactions to the type of SSM was very weak. In previous
calculation we used the parameters of the BS2005(OP) model
presented on website \cite{bahcall95}. For the case of the
exponential potential we use additionally BP2000 Solar Standard
Model \cite{bahcall95} to check sensitivity of the flux to the
type of  SSM.

\section{The probability of the $\mathbf{pep\to d+\nu}$ reaction}
\label{gen}
It should be noted that Bahcall and May \cite{bahcall69} used the
adiabatic approximation for the wave function of the $pep$ system.
In this approach the $pep$ wave function was presented by the
product of the wave function of electron moving relative to the
center of mass of the two protons and the wave function of the
relative motion of two protons. Unfortunately, such a description
is not adequate even in the region where the distance from the
center of mass of the protons to electron far exceeds the size of
nucleon-nucleon system due to the nature of the Coulombic force
\cite{alt93}. Moreover, we cannot use the such factorized wave
function at the region of the small distances between the
particles where it is necessary to know the wave function with a
sufficient accuracy to perform a precise calculation of the
transition matrix element of the process $pep\to d\nu_e$.

In our paper \cite{bah13} we showed the use of hyperspherical
harmonics method \cite{djibuti,fabre}  to directly solve the
3-body Schr\"odinger equation for $pep$ system. Deduction of
one-dimensional coupled radial equations can be found in our
article \cite{bah13}. Due to the smallness energy of the colliding
particles we restrict ourselves by consideration of the radial
equation assuming that all quantum numbers are zero. Therefore we
solve the following radial equation:
\begin{equation}\label{rad eq0}
\frac{d^2U(\rho)}{d\rho^2}+\frac{1}{\rho}\frac{dU}{d\rho}-\Bigl({\cal{V}}(\rho)+\frac{4}{\rho^2}-\kappa^2\Bigr)U(\rho)=0,
\end{equation}
where $\kappa^2=2\mu_{23}E/\hbar^2>0$ ($E$ is the total energy of
the colliding particles in the $pep$ system);
\begin{eqnarray}\label{Vpot0}
{\cal{V}}(\rho)\,\,\,\,&=&{\cal{V}}^N(\rho)+{\cal{V}}^C(\rho),\\
{\cal{V}}^C(\rho)&=&\frac{32\mu_{23}}{3\pi\hbar^2}\frac{(a_1+a_2+a_3)}{\rho}\equiv\frac{2\eta_3\kappa}{\rho}.
\end{eqnarray}
Here $\eta_3$ is the 3-body Coulomb parameter (analogue of the
Sommerfeld parameter) which is defined as
\begin{eqnarray}\label{eta}
\eta_3&=&\frac{16\mu_{23}}{3\pi\hbar^2\kappa}(a_1+a_2+a_3),\\
a_1&=&\sqrt{\frac{m_2m_3}{\mu_{23}(m_2+m_3)}}e^2\equiv e^2,\label{a1}\\
a_2&=&-\sqrt{\frac{m_1m_3}{\mu_{23}(m_1+m_3)}}e^2\simeq
-\sqrt{\frac{2m_1}{m_3}}e^2,\label{a2}\\
a_3&=&-\sqrt{\frac{m_1m_2}{\mu_{23}(m_1+m_2)}}e^2\simeq
-\sqrt{\frac{2m_1}{m_2}}e^2.\label{a3}
\end{eqnarray}
${\cal{V}}^N$ and ${\cal{V}}^C(\rho)$ are the reduced nuclear and
Columbic potentials resulting due to integration by applying the
hyperspherical function of zero value of the hypermoment. These
potential can be easier calculated analytically for the considered
potentials.

To find numerically a solution of Eq. (\ref{rad eq0}) we use the
following boundary condition at the point $\rho_0$ closed to the
origin:
\begin{equation}\label{b1.con}
U(\rho_0)=J_2(\kappa_0 \rho_0),\,\,\,\, U^\prime (\rho_0)=\kappa_0
J_2^\prime(\kappa_0 \rho_0),
\end{equation}
in the case ${\cal{V}}^N(\rho_0)<0$ (pure attractive potential)
or
\begin{equation}\label{b2.con}
U(\rho_0)=I_2(\kappa_0 \rho_0),\,\,\,\, U^\prime (\rho_0)=\kappa_0
I_2^\prime(\kappa_0 \rho_0),
\end{equation}
in the case ${\cal{V}}^N(\rho_0)>0$ ($NN$ potential is repulsive
near the origin). Here
$\kappa_0=\sqrt{\kappa^2+\mid{\cal{V}}^N(\rho_0)\mid}$
$(\kappa_0=\sqrt{{\cal{V}}^N(\rho_0)-\kappa^2}$ if
${\cal{V}}^N(\rho_0)>0$), $J_2(x)$ is the Bessel function and
$I_2(x)$ is the modified Bessel function. Such boundary condition
follows from behavior of the radial function near origin. It is
clear to see that $U(\rho)\sim \rho^2$ if $\rho\to 0$ for any type
of $NN$ potential without hard core.

The asymptotic of the radial function $U(\rho)$ at large distance
($\rho \rightarrow \infty$) is:
\begin{equation}\label{assCoul}
U(\rho)\longrightarrow e^{i\delta_3}\cos\delta_3\bigl(F_{00}(\kappa\rho)-\tan\delta_3G_{00}(\kappa\rho)\bigr),
\end{equation}
where $\delta_3$ is the 3-body nuclear scattering  phase shift
modified by the Coulomb interactions. At the considered low energy
range the phase $\delta_3\sim 0$  and according to Ref.
\cite{djibuti} the 3-body Coulombic phase shift is given by
\begin{equation}
\delta_{3C}=\arg\left[\Gamma(5/2+i\eta_3)\right],
\end{equation}
and  $F_{00}(\kappa\rho)$ and $G_{00}(\kappa\rho)$ are the 3-body
regular  and irregular Coulomb wave functions \cite{djibuti}. A
reader can find the representations of these functions  in our paper
\cite{bah13}.

Eq. (\ref{rad eq0}) was solved using the boundary conditions
(\ref{b1.con}) or (\ref{b2.con}) depending on the behavior of the
$NN$ potential near origin. Equating the logarithmic derivative of
the numerical solution at distance where the nuclear potential is
negligible to the logarithmic derivative of the asymptotic
solution [Eq. (\ref{assCoul})] we find the 3-body phase shift
$\delta_3$. The result for $U(\rho)$ obtained with the Gauss
potential was shown in Figure 1 of Ref. \cite{bah13}. We got the
same results with the Yukawa and exponential potentials. The limit
of $U(\rho)/(\kappa\rho)^2$ goes to nonzero value. However, near
the origin, the behavior of the curve describing solution with the
M-T potential differs from the solution applying simple
potentials. In this case the limit of
$U(\kappa\rho)/(\kappa\rho)^2$ approaches to the value close to
zero due to the repulsive term in the M-T potential which becomes
very large at small distance. However, the behavior of radial
function is almost the same at large distance like for the
simplest potentials having only attraction term. At sufficiently
large distances, the ratio of the unnormalized solution of the
equation of Eq.(\ref{rad eq0}) to the asymptotic function given by
Eq. (\ref{assCoul}) becomes constant for all considered cases. It
allowed us to get the normalized wave function \cite{bah13}. The
matching distance depends on the type of potential and lies at a
distance $\sim$35 fm for the case of the simple $NN$ potentials
and in the case of M-T potential matching distance is $\sim$50 fm
owing to its repulsive part of interaction. It should be noted
that the wave function of $pp$ system can be replaced by its
asymptotics already from a distance of $\sim$5 fm. For this
reason, the rate of the $pp$ reaction is insensitive to the choice
of potential. At the same time in the calculation of the $pep$
reaction the wave function of $pep$ system can be replaced by its
asymptotics from a distance of only $\sim$35-50 fm, but for the
distance less than 35 fm we must use the exact 3-body wave
function.  The 3-body wave function must be calculated with
sufficient accuracy at range $0< \rho < 35$ fm because the
deuteron wave function decays exponentially to zero around 35 fm.
Interested readers can study dependence of the $pep$ 3-body wave
function on the hyperradius in Figure 1 of our paper \cite{bah13}.

Applying the weak Hamiltonian (Eq. (\ref{Hweak})) the matrix
element for the $pep\to d\nu$ transition can be written as
\begin{eqnarray}\label{matrix}
H_{if}&=& G_A
<\varphi_\nu\mid{\mathbf{\sigma}}\tau^{(+)}\mid\varphi_e>\times\nonumber\\
&&\sum\limits_{i=1}^{2}<\Psi_d\Psi_\nu\mid{\mathbf{\sigma}}_i\tau_i^{(-)}\mid\Psi_{pep}>,
\end{eqnarray}
where  $\varphi_\nu$ and $\varphi_e$ refer the spin functions of
neutrino and electron, respectively; the wave function of the
deuteron is denoted by $\Psi_d$, and $\Psi_{pep}$ is the $pep$
wave function. We take the neutrino wave function $\Psi_\nu$ as a
plane wave which can be replaced by unity owing to cutoff of the
integration interval in the calculation of the matrix element (Eq.
(\ref{matrix})). Due to the zero-range weak interaction of
electron with proton the electron coordinate in the wave function
$\Psi_{pep}$  is taken at the point where either of the protons is
located.

The probability $P_3$ of reaction per unit time was calculated
using the first-order perturbation theory in the weak interaction.
We define the probability $P_3$ as
\begin{equation}\label{prob}
P_3=\frac{2\pi}{\hbar}\overline{\mid H_w\mid^2}\rho(E_\nu),
\end{equation}
where  the density of neutrino states is
\begin{equation}
\rho(E_\nu)=\frac{E_\nu^2}{2\pi^2\hbar^3c^3}.
\end{equation}
Here $E_\nu=$1.44 MeV is the neutrino energy and $c$ is light
speed.

Finally, we get the equation for the probability $P_3$ of the
$pep\to d\nu$ reaction in form:
\begin{equation} \label{P_3}
P_3=\frac{3E_\nu^2G_A^2}{\pi\hbar^4c^3}\mid\int\Psi^*_d({\mathbf{x}})
\Psi_{pep}({\mathbf{x,y_0}})d^3x\mid^2,
\end{equation}
where ${\mathbf{y_0}}$ is a Jacobi coordinate of the electron at
the point where it contacts with one of the protons.

\begin{figure}
\epsfig{file=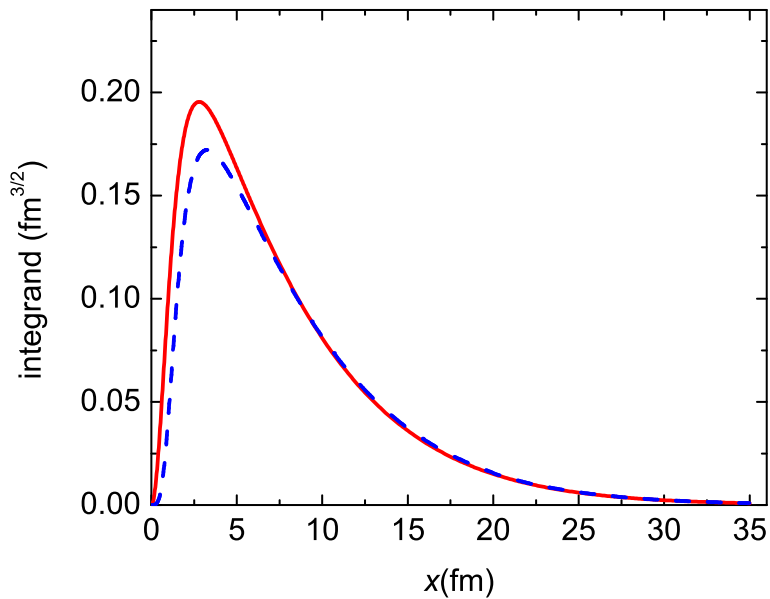,width=12.0cm} \caption{ The dependence
of the integrand in Eq. (\ref{overlap}) on the distance calculated
with the exponential  (solid line) and M-T potentials (dashed
line) at $E_{pep}$=8 keV. \label{fig1}}
\end{figure}

The overlap integral of  Eq. (\ref{P_3}) can be reduced to the one
dimensional integral:
\begin{eqnarray}\label{overlap}
\int\Psi_d^*({\mathbf{x}})\Psi_{pep}({\mathbf{x,y_0}})d^3x=
\frac{16\sqrt{\pi}}{\kappa^2}\int\limits_0^\infty
x^{-1}\chi_d(x)U(x)dx.
\end{eqnarray}
Here the function $U(\rho)$ is taken at the point
$\rho=\sqrt{x^2+y_0^2}\simeq x$ ($x\gg y_0$) and the radial wave
function $\chi_d(x)$ of the deuteron is normalized to unity.
Figure \ref{fig1} shows the results of the dependence of the
integrand of the matrix element (\ref{overlap}) calculated with
the exponential and M-T potentials. We see that the integrands
depend on whether the $NN$ potential has repulsion at small
distances or whether it is a purely attractive potential.

In nuclear astrophysics the rate of the binary reaction is
calculated using the astrophysical $S$ factor \cite{angulo}. The
cross section of such reactions is related to the astrophysical
$S$ factor as
\begin{equation}\label{S_pp}
\sigma(E)=\frac{S(E)}{E}e^{-2\pi\eta},
\end{equation}
where $\eta=Z_1Z_2e^2/(\hbar v)$ is the Sommerfeld parameter,
$Z_1e$ and $Z_2e$ are charges of colliding nucleus with the
relative velocity $v$. The factor $\exp(-2\pi\eta)$ is
proportional to the probability of the penetration of a charged
particle through the Coulomb barrier. In the $pep$ reaction we can
adopt a similar parameter because the 3-body $pep$ radial wave
function $U(\rho)$ encloses the factor
$\exp(-\pi\eta_3/2)\mid\Gamma(5/2 + i\eta_3)\mid$ owing to the
Coulomb interaction between colliding particles. Therefore we can
identify the Coulombic factor in the probability $P_3$ and define
the astrophysical $S_{pep}$ factor for the $pep$ reaction as
\begin{eqnarray}\label{S_3}
P_3(E)&=&G_0(E)S_{pep}(E),\label{gamow_m}\\
G_0(E)&=&\frac{2\pi
e^{-2\pi\eta_3}(\frac{1}{4}+\eta_3^2)(\frac{9}{4}+\eta_3^2)}{1+e^{-2\pi\eta_3}}.\label{gamow_3}
\end{eqnarray}
For range of energy close to zero, our results show that
$S_{pep}(E)$ varies almost linearly with energy $E$. Its limit is
not zero when the energy goes to zero. Also we note that $G_0(E)$
is dimensionless therefore the $S_{pep}(E)$ dimension coincides
with the unit of $P_3(E)$ according to our definition of this
quantity. Due to linear property of $S_{pep}(E)$ at low energies
we can expand it into a series and restrict ourselves to first
three terms, so we write
\begin{equation}\label{expand}
S_{pep}(E)=S_0+S_1 E+S_2E^2,
\end{equation}
The calculated results for the value of the
coefficients are following:\\
the exponential potentials
\begin{eqnarray}\label{coef-expon}
S_0&=&2.49\times 10^{10} \rm{fm^6/s},\nonumber\\
 S_1&=&3.35\times
10^{10}
\rm{fm^6/(MeV\,s)}, \nonumber\\
S_2&=&2.21\times 10^{10} \rm{fm^6/(MeV^2\, s)};
\end{eqnarray}
and the M-T potentials
\begin{eqnarray}\label{coef-M-T}
S_0&=&2.11\times 10^{10} \rm{fm^6/s},\nonumber\\
 S_1&=&3.70\times
10^{10}
\rm{fm^6/(MeV\,s)}, \nonumber\\
S_2&=&1.76\times 10^{10} \rm{fm^6/(MeV^2\, s)}.
\end{eqnarray}
The coefficients $S_0$, $S_1$ and $S_2$ for the Gauss and the
Yukawa potential are given in Ref. \cite{bah13}. We note the
difference between values of the coefficients $S_0$ for considered
simple potentials (Gauss, Yukawa, exponential) can reach
$\sim$3-6\%, while the difference for the coefficient $S_0$
calculated with M-T potentials reaches $\sim$15\% with respect to
the simplest potentials. The difference for the coefficients $S_1$
and $S_2$ goes to  much larger values. Fortunately the
contribution of $S_1$ and $S_2$ to the flux of neutrino is much
less as compared to $S_0$. We would also like to mention that the
analogue astrophysical $S$ factor for two particle reaction
(\ref{pp}) does not depend on the type of $NN$ potential in
contrast to the three particle astrophysical $S$ factor defined by
Eq. (\ref{gamow_3}).

\section{Rate and flux of the solar neutrinos}
\label{rflux}
We define  the rate constant of the $pep$ reaction
\cite{bah13}  by following equation
\begin{eqnarray}\label{rate}
{\cal{K}}_{pep}=<P_3>=
\int\limits_0^\infty\int\limits_0^\infty\int\limits_0^\infty\varphi(v_e)\varphi(v_{p_1})\varphi(v_{p_2})P_3(E)dv_edv_{p_1}dv_{p_2}.
\end{eqnarray}
After averaging over the Maxwell-Boltzmann distributions
$\varphi(v_i)$ describing the random motion of protons and
electrons in the core plasma of the Sun we get the simple formula
for the rate constant expressed though the 3-body astrophysical
$S_{pep}$ factor:
\begin{equation}\label{r-k}
{\cal{K}}_{pep}=\frac{1}{(kT)^{3}}\int\limits_0^\infty G_0(E)
S_{pep}(E) e^{-E/kT}E^2dE,
\end{equation}
where $k$ is the Boltzmann constant and $T$ is temperature in the
core of the Sun. The maximum of the integrand is reached at the
energy
\begin{equation}\label{max-E}
E_{max}=\Bigl(\frac{1}{2}kT\sqrt{E_G}\Bigr)^{2/3}.
\end{equation}
Here $E_G$ is the Gamow  energy for $pep$ reaction \cite{bah13}.

The rate of reaction related to the rate constant is
\begin{equation}\label{rate-n}
{\cal{R}}_{pep}={\cal{K}}_{pep}n_p^2n_e,
\end{equation}
where $n_e$ and $n_p$ are the density of the reactants (protons
and electrons).

In our calculation of the $pp$ and $pep$ reaction rates we applied
the BS2005(OP) SSM \cite{www-sns-ias-edu} for both type of $NN$
potentials. The $pp$ and $pep$ reaction rates do not depend much
on the type of the solar model (for details we refer to
\cite{bahcall-sab2}). Additionally we used BP2000 SSM for
calculation of the rate with the exponential potentials to check
sensitivity of the results to the SSMs.

\begin{figure}
\epsfig{file=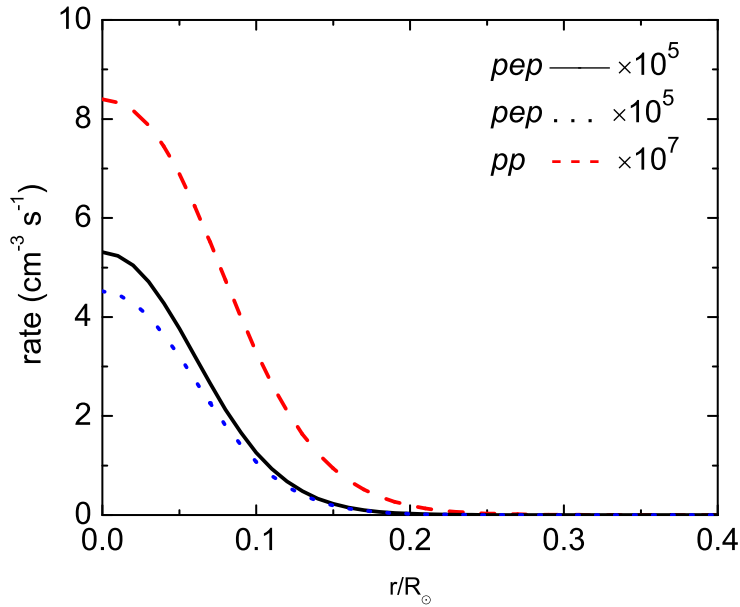,width=12.0cm} \caption{The $pep$ (solid
line for exponential potential  and dotted line for M-T potential)
and $pp$ (dashed line for exponential potential) rates vs. the solar
interior. The results of the $pp$ rate is almost same for the
exponential and M-T potentials.
BS2005 (OP) SSM is
applied. \label{fig3}}
\end{figure}

In  Figure {\ref{fig3}} we present the dependence of the rate of
the $pep$ and $pp$ reactions on the solar interior calculated by
the exponential and M-T potentials. The rate of $pp$ reaction
differs within 2.4\% for all type of used potentials, including
the potential used in Ref. \cite{bah13}. Therefore in the figure
the rate of $pp$ reaction is shown only for the exponential
potential. There is a small difference of the rate of $pep$
reaction calculated by applying the simplest attractive
potentials. However, in the case of M-T potential it differs
within $\sim$10-15\% from the rates calculated with the simplest
potentials.

Finally we present, in Table \ref{tab1}, the results of
calculation of the fluxes of neutrinos at a distance of one
astronomical unit (AU).
\begin{table}[!t]
\caption{Predicted fluxes $\Phi_{pp}$ and $\Phi_{pep}$ (without
survival probability), in units of $10^{10}\,(pp), 10^8\,(pep)
\,\,\rm{cm^{-2}s^{-1}}$.}
\begin{tabular}{lcccl} \hline
 Standard Solar&\,$\Phi_{pp}$\,
&\,$\Phi_{pep}$\,&$\Phi_{pp}/\Phi_{pep}$&
   References\\
    Model & &&&\\
 BS2005(OP)&6.20 &2.04&304 & using Gauss potential  \cite{bah13}\\
 &6.05 &1.99&304 & using Yukawa potential  \cite{bah13}\\
 &6.13 &2.14&287 & using exponential potential  \\
 &&&&(current work)\\
 &6.16 &1.82&338 & using M-T potential (current work)\\
 BP2000&6.02 &2.10&288 & using exponential potential\\
 &&&&(current work)\\
 BP04(Yale)&5.94&1.40&424 & \,\,\,\, \cite{bahcall-sab1} \\
 BP04(Garching)&5.94&1.41&421 &\,\,\,\, \cite{bahcall-sab1} \\
 BS04&5.94&1.40&424 & \,\,\,\, \cite{bahcall-sab1} \\
 BS05(14N)&5.99&1.42&421 & \,\,\,\, \cite{bahcall-sab1} \\
 BS05(OP)&5.99&1.42&421 & \,\,\,\, \cite{bahcall-sab1} \\
 BS05(AGS,OP)&6.06&1.45&418 & \,\,\,\, \cite{bahcall-sab1} \\
 BS05(AGS,OPAL)&6.05&1.45&417 &\,\,\,\, \cite{bahcall-sab1}  \\
 \hline
 \end{tabular}
\label{tab1}
\end{table}
One notes that whereas our results of calculated flux of neutrino
for $pp$ reaction are close to the Bahcall \textit{et al.} results
but  differences exist for the calculated results of the $pep$
reactions.

If we take into account the Borexino collaboration measured
neutrino flux ($\Phi_{pep}=(1.6\pm 0.3)\times 10^8
\,{\rm{cm}}^{-2} {\rm{s}}^{-1}$) and  the survival probability of
the electron neutrino ($P=0.62 \pm 0.17$ at 1.44 MeV) suggested by
this group \cite{borexino} we find that the neutrino flux at 1 AU
should be equal to $\Phi_{pep}= (1.33\pm 0.36)\times 10^8\,\,
{\rm{cm}}^{-2}{\rm{s}}^{-1}$ for the exponential potential, and
$\Phi_{pep}=(1.13\pm 0.31)\times 10^8\,\,
{\rm{cm}}^{-2}{\rm{s}}^{-1}$ for the M-T potential if the
BS2005(OP) Standard Solar Model is used. Applying BP2000 Standard
Solar model leads to the flux equal to $\Phi_{pep}= (1.30\pm
0.36)\times 10^8\,\, {\rm{cm}}^{-2}{\rm{s}}^{-1}$ in the case of
the exponential potential.

We see that our calculated neutrino fluxes from the $pep$ reaction
lie within the confidence interval of the experimental data for
all considered simplest  potentials including the exponential
potential even if we multiply the calculated fluxes by the
survival probability ($P=0.62 \pm 0.17$) suggested by Borexio
group. The flux calculated by using M-T potential lies at the
lower limit of the confidence interval. If we multiply the fluxes
of neutrino from $pep$ reaction calculated by Bahcall \textit{et
al.} to the value of survival probability  all  their results  lie
out of the confidence interval of the Borexino data.

Averaging the Bahcall \textit{et al.} fluxes from $pp$ reaction
presented in Table \ref{tab1} give us the value equal to
$\Phi_{pp}=5.99\times 10^{10}\, {\rm{cm}}^{-2} {\rm{s}}^{-1}$ and
with standard deviation equal to $0.05 \times 10^{10}\,
{\rm{cm}}^{-2} {\rm{s}}^{-1}$.  Comparing the fitted low-energy
parameters for the simplest potentials, we see that the difference
between effective range parameters may be  2\%-7\%, at the same
time the neutrino fluxes from the $pp$ reaction have a 3.5\%
maximal difference from the average value of  Bahcall result (in
the case of the Gauss potential, BS2005 (OP) SSM). This means the
fluxes from $pp$ reaction is insensitive  to the type of $NN$
potential and to SSM. However, the difference between the fluxes
of neutrinos from $pep$ reaction calculated by the simplest
potentials and the potential having repulsion at small distances
(M-T potential) may reach $\sim 15\%$. Consequently our results
from $pep$ reaction is expected to show dependence on the type of
used $NN$ potential. Also we must note that the difference between
our results and Bahcall \textit{et al.} results for $pep$ process
can reach up to 25\% to 40\%, and indicates  strong dependence on
the selection of the initial 3-body state of the $pep$ system.

\section{Conclusion}\label{fin}
The process $pep\to d + \nu_e$  has been considered for the first
time within the framework of the 3-body description of the initial
state employing the nucleon-nucleon potential having repulsion at
small distances. The rate and the flux of neutrino from $pp$ and
$pep$ reactions has been calculated for specific conditions in the
center of the Sun. The theoretical results are in satisfactory
agreement with the reported measured data by the Borexino group,
but different by (25-40)\% from previous calculations made by
Bahcall \textit{et al.} For a better comparison between theory and
measurement it is necessary to reduce uncertainty in oscillation
parameters of oscillation of neutrinos and to make a new
calculation using other realistic nucleon-nucleon potentials
having repulsive core.

\end{document}